\documentstyle[12pt,epsfig,rotating]{article}
\newcommand{\AmS}{{\protect\the\textfont2
  A\kern-.1667em\lower.5ex\hbox{M}\kern-.125emS}}
\def\E{\mbox{e}^+\mbox{e}^-}

\def\ifmath#1{\relax\ifmmode #1\else $#1$\fi}%

 \def\max{\ifmath{{\mathrm{max}}}}
\def\min{\ifmath{{\mathrm{min}}}} 
\def\s2{\hskip2pt}      
   
\newcommand{\beqa}{\begin{eqnarray}} \newcommand{\eeqa}{\end{eqnarray}  }
\newcommand{\beqan}{\begin{eqnarray*}} \newcommand{\eeqan}{\end{eqnarray*}}
\newcommand{\beq}{\begin{equation}} \newcommand{\eeq}{\end{equation}  }
\def\cl{\centerline} \def\bcc{\begin {center}} \def\ecc{\end {center}}
\def\btbl{\begin{tabular}} \def\etbl{\end{tabular}}
 \def\NUFM{{\footnotesize NUFM}}
 
 \def\SPS{{\footnotesize SPS}}
\def\E877{{\footnotesize E877}} 
 \def\AGS{{\footnotesize AGS}}
\def\RHIC{{\footnotesize RHIC}} 

\def\E917{{\footnotesize E917}}
\def\BRAHMS{{\footnotesize BRAHMS}}
\begin{document}



\centerline{\bf Collective Flow Distributions and Nuclear Stopping
} \vskip0.3cm \centerline{\bf in Heavy-ion Collisions at AGS, SPS
and RHIC}

\vskip1cm \centerline{ Shengqin Feng, Xianbao Yuan and Yafei Shi}
\vskip0.4cm \centerline{\small Dept of Physics, Science College,
China Three Gorges University, Yichang, 443002, China}

\vskip0.9cm

\vskip2cm
\begin{center}\begin{minipage}{124mm}
{\small \hskip0.8cm We study the production of proton, antiproton
and net-proton at \AGS, \SPS~ and \RHIC~ within the framework
non-uniform flow model(\NUFM) in this paper. It is found that the
system of \RHIC~ has stronger longitudinally non-uniform feature
than \AGS~ and \SPS~, which means that nuclei at \RHIC~ energy
region is much more transparent. The \NUFM~ model provides a very
good description of all proton rapidity at whole \AGS~, \SPS~ and
~\RHIC. It is shown that our analysis relates closely to the study
of nuclear stopping and longitudinally non-uniform flow
distribution of experiment. This comparison with \AGS~ and \SPS~
help us to understand the feature of particle stopping of thermal
freeze-out at \RHIC~ experiment.}
\end{minipage}\end{center}
\vskip3cm PACS number(s): {25.75.-q }

\newpage

\begin{center}{\large \bf I.~~Introduction}\end{center}

\noindent Nuclear matter is believed to be compressed to high
baryon density during central collisions~[1,~2,~3]. In the
interaction region of the colliding nuclei, nucleons undergo
collisions which reduce their original longitudinal momentum. This
loss of rapidity is an important characteristic of the reaction
mechanism and is often referred to as stopping power~[4].
Relativistic heavy-ion collisions are unique in the sense that
secondary collisions of excited baryons are expected to contribute
to rapidity loss leading to simultaneous stopping of many nucleons
within the interaction volume.

Proton and antiproton provide an experimental tool for measuring
baryon production and allow us to explore baryon transport from
beam rapidity toward mid-rapidity [5,~6]. The global thermodynamic
properties and collective motion of the system at the kinetic
freeze-out point can be deduced, albeit in a model-dependent way.
\BRAHMS\ collaboration [5] working at \RHIC~ has published
rapidity distributions of net proton at central collisions. These
results will help us to study the dynamic feature of \RHIC~ and
collective movement(flow).

The study of collective flow in high energy nuclear collisions has
attracted increasing attentions from both experimental and
theoretical point of views. The rich physics of longitudinal and
transverse flows is due to their system evolution at early time
and nuclear stopping. In general, the collective evolution of the
hot and dense matter leaves a distinct imprint on the phase space
distribution of the fireball at freeze-out. To disentangle such
information from features generated during freeze-out a refined
understanding of the decoupling process is needed.

Here we should mention three kinds of models of thermal and
collective flow, the first one is the spherically-expanding source
may be expected to approximate the fireball created in
lower-energy collisions.

At higher energies stronger longitudinal may lead to a cylindrical
geometry according to the second kind model postulated by
Schnedermann, Sollfrank and Heinz~[7]. It accounted for the
anisotropy of longitudinal and transverse direction by adding the
contribution from a set of fireballs with centers located
uniformly in the rapidity region in the longitudinal direction
which sketched schematically in Fig.1 and 2 of Ref.[8]. It can
account for the wider rapidity distribution when comparing to the
prediction of pure thermal isotropic model.

The third kind of model is postulated by Bjorken~[9] which has
been formulated for asymptotically high energies, where the
rapidity distribution of produced particles establishes a plateau
at mid-rapidity. As mentioned before[5-10], collisions at
available heavy-ion energy regions of ~\AGS, \SPS~ and \RHIC~ are
neither fully stopped nor fully transparent, although a
significant degree of transparency are observed. Consequently the
overall $dN/dy$ distribution of baryons is expected to consist of
the sum of the particles produced in the boost-invariant central
zone and the particles produced by the excited fragments. The fact
that the observed distributions are flatter at mid-rapidity and
wider than those predicted by the thermal isotropic model might
point in this direction. Especially at the \SPS~ and \RHIC~ energy
region, the central dip of baryon rapidity distribution is clearly
shown by the experiment. A nonuniform longitudinal flow model was
proposed~[8,~10], which assumed that the centers of fireballs were
distributed non-uniformly in the longitudinal phase space.

This paper is organized as follows. In Sec.II we give a
qualitative description with a geometrical parametrization picture
and briefly review the longitudinal Non-Uniform Flow
Model(\NUFM~)~[8]. The comparison and analysis of baryon
distribution of \AGS,~\SPS, and \RHIC~ with the calculation
results of the model are given in Sec.III. Sec.IV contains a
summary and conclusions.

\vskip1.0cm

\begin{center}{\bf \large II.~The main idea of Non-Uniform Flow
Model(~NUFM~)}\end{center}

\noindent The \NUFM~~ model we considered [~8] contains three
distinct assumptions some of which are rather different from those
usually contained in other fireball models:

(1) We argue that the transparency/stopping of relativistic heavy
ion collisions should be taken into account more carefully. A more
reasonable assumption is that the fireballs keep some memory on
the motion of the incident nuclei, and therefore the distribution
of fireballs, instead of being uniform in the longitudinal
direction,is more concentrated in the direction of motion of the
incident nuclei,i.e.,more dense at large absolute value of
rapidity. It will not only lead to anisotropy in
longitudinal-transverse directions, but also render the
fireballs(especially for those baryons) distribute non-uniformly
in the longitudinal direction shown in Fig.1.

(2) We assume that whether it is at higher ~\RHIC~ energy region
or lower ~\AGS~ energy region, the freeze-out temperatures are
nearly same, around 120 MeV. Since the temperature at freeze-out
exceeds 100MeV, the Boltzmann approximation seems reasonable to
study ~\RHIC~ energy region at freeze-out.

(3) An ellipticity parameter $e$ is introduced through a
geometrical parametrization which can express the non-uniformity
of flow in the longitudinal direction. For the central collisions,
the nuclear stopping  can be studied  by the range of rapidity of
emission source ($-y_{\rm e0}<y<y_{\rm e0}$) in the center-of-mass
system ($y=y_{\rm cm}$).

A parametrization for such a non-uniform distribution can be
obtained by using an ellipse like picture on emission angle
distribution, as shown in Fig.2, in this scenario, the emission
angle is:

\noindent
\begin{equation} 
\vartheta=\textrm{tan}^{-1}(e\textrm{tan}\Theta)
\end{equation}

Here, the induced parameter $e$($0\leq{e}\leq1$) represents the
ellipticity of the introduced ellipse which describes the
non-uniform of fire-ball distribution in the longitudinal
distribution, The detailed discussions of the \NUFM~ was given at
the Ref.[8]. The rapidity distribution of \NUFM~ is:

\begin{equation} 
\frac{dn_{NUFM}}{dy}= eKm^{2}T\int_{\kappa_\min}^{\kappa_\max}
\frac{Q(\vartheta)d\vartheta}{sin(\vartheta)}(1+2\Gamma+2\Gamma^{2})e^{(-1/\Gamma)}
\label{eq:dndy1}
\end{equation}

Here $K$ is a constant, $\kappa_{\min}=2\tan^{-1}(e^{-y_{e0}})$,
$\kappa_{\max}=2\tan^{-1}(\textrm{e}^{y_{e0}})$, $y_{e0}$ is the
rapidity limit which confines the rapidity interval of
longitudinal flow. In Eq.2,

\begin{equation} 
\Gamma=T/{m\cosh(y-y_{e})},
\end{equation}

\begin{equation} 
Q(\vartheta)=\frac{1}{\sqrt{e^{2}+\tan^{2}\kappa}\mid\cos\vartheta\mid},
\end{equation}

Here $\kappa=2\tan^{-1}(e^{-y_{e}})$, $y_{e}$ is the rapidity of
collective flow, $m$ is the mass of produced particle, $T$ is the
temperature parameter. The physical meanings of parameters $e$ and
$y_{e0}$ will be discussed in the paper. Changing the integration
variable in Eq.2 back to $y_{e}$, the rapidity distribution can be
rewritten aa follows:

\begin{equation} 
\frac{dn_{NUFM}}{dy}=eKm^{2}T\int_{-y_{eo}}^{y_{eo}}\rho(y_{e})
dy_{e}(1+2\Gamma+2\Gamma^{2})e^{(-1/\Gamma)},
\end{equation}

Here
$\rho(_{e})=\sqrt{\frac{1+\sinh^{2}(y_{e})}{1+e^{2}\sinh^{2}(y_{e})}}$
is the flow distribution of longitudinal flow, $e$ is the
parameter ($0\leq e\leq1$) represents the ellipticity of the
introduced ellipse which describes the non-uniform of fire-ball
distribution in the longitudinal direction, as sketched in Fig.2
It can be seen from Fig.3 that the larger is the parameter $e$,
the flatter is the distribution of $\rho(_{e})$ and the more
uniform is the longitudinal flow distribution. When
$e\rightarrow1$, the longitudinal flow distribution is completely
uniform $\rho(_{e})\rightarrow1$.

\vskip1.0cm

\begin{center}{\bf \large III Comparison with experiments}\end{center}

\vskip0.8cm
\begin{center}{\bf A. Rapidity distributions of protons at
AGS}\end{center}

In this part, we will discuss the mid-rapidity baryon productions
at \AGS~. This is because rapidity distributions are strongly
affected by "memories of the pre-collision state": Whereas all
transverse momentum are generated by the collision itself, a
largely unknown fraction of the beam-component of the momenta of
the produced hadrons is due to the initial longitudinal motion of
the colliding nuclei. In hydrodynamics one finds that final
rapidity distributions are very sensitive to the initialization
along the beam direction, and that hydrodynamics evolution is not
very efficient in changing the initial distributions~[11]. The
best way to isolate oneself longitudinally from remnants of the
initially colliding nuclei is by going as far away as possible
from the projectile and target rapidities, i.e. by studying
mid-rapidity hadron production.

The rapidity distributions of proton for beam kinetic energies of
6,8,and 10.8GeV/n Au + Au collisions by \E917~[3,12,13] are given
in Fig.4. The dotted and solid lines correspond to the results
from isotropic thermal model and nonuniform longitudinal flow
model(\NUFM), respectively. The rapidity limit $y_{\rm e0}$ and
ellipticity $e$ at different collision energies used in Table I
for comparison. The parameter $T$ is chosen to be 0.12 GeV.

\vskip0.5cm \cl{Table I \ \ The value of model-parameters at AGS}
\vskip-0.5cm \bcc\btbl{|c|c|c|c|}\hline
           & \multicolumn{3}{|c|}{Au-Au Collisions energy GeV/n}
        \\ \cline{2-4}
 Parameter & 6 & 8 & 10.8
\\ \hline
$e$ & 0.901 & 0.852 & 0.802
\\ \hline
$y_{\rm e0}$ & 1.213 & 1.263 & 1.312
\\ \hline \etbl\ecc

Fig.4 shows the $dN/dy$ distributions for the most central event
classes at all three energies plotted as open circle points of
experimental data. the dotted curves show the expected
distribution for completed stopping i.e.,isotropic emission
protons with a Boltzmann energy distribution. It completely fails
for the $dN/dy$ distributions. The measured rapidity distributions
at all three beam energies suggested that the degree of stopping
in central collisions is not completely stopping  at the \AGS~
energy region. And the non-uniformity parameter $e<1$ also
suggests that the collective flow in the longitudinal direction at
\AGS~ are not completely uniform.

In Ref.8, we compared the parameter values for Si+Al(smaller
colliding nuclei) and Au+Au (larger colliding nuclei) collisions
listed in Table I of Ref.8. It shown that there was stronger
nuclear stopping in the collision of larger nuclei. Here we
compare with different collision energies with same collision
system. It can be  shown from Table.I that the rapidity limit
$y_{\rm e0}$ is bigger and the ellipticity $e$ for is smaller for
higher collision energy at \AGS, which means that decreasing
nuclear stopping and uniformity in the longitudinal direction with
increasing incident energy.

\vskip0.8cm
\begin{center}{\bf B. Rapidity distributions of net-proton at different regions}\end{center}

\noindent Fig.5 shows net-proton $dN/dy$ measured at
\AGS[3,14,15], \SPS~[16] and \RHIC~[5] from the top $5\%$ central
collisions. the real triangle, open square and open circle points
are for the \AGS~ ,\SPS~ and \RHIC~ experiment results,
respectively. The solid lines are our \NUFM~ calculation results.
It can be seen from the Fig.5 that \NUFM~ model reproduces the
central dip of the rapidity distribution of the proton at \SPS~
and \RHIC~ in agreement with the experimental findings. Note that
the appearance or disappearance of central dip is insensitive to
the rapidity limit $y_{\rm eo}$, but depends strongly on the
magnitude of the ellipticity $e$ for proton distribution. From
that we can see that the central dip  strongly relates with the
non-uniformity of longitudinal flow distributions. On the other
hand, The second parameter in \NUFM\ $y_{\rm eo}$ determines the
width of the rapidity distributions. The larger of the
distribution limit of emission source $y_{\rm eo}$ , the wider of
the rapidity distributions. For \SPS\ and \RHIC\, $y_{\rm eo}$ are
approximately equal to the half width of two peak distribution. In
the sense, the parameter $y_{\rm eo}$ can represent the nuclear
stopping power. The parameter of $T$ is chosen to be 0.12 GeV.

The distributions show a strong energy dependence: at \AGS, the
net protons distribution has a peak at mid-rapidity, the
distribution at \AGS~ is narrower than the other two energies, and
$e=0.7327$ which shows no dip at the central rapidity, the
collective flow is approximately uniform and the particle
distribution shows a peak at the central of rapidity. While at
\SPS\,  a dip is observed in the middle of rapidity of the
distribution. As comparison with \AGS~, as the decreasing of $e$,
the collective flow at $y\sim 0$ is diluted at \SPS~. At \RHIC~ a
broad minimum (dip) has developed spanning several units of
rapidity, indicating that at \RHIC~ energies collisions are quite
transparent. According to the study of ellipticity $e$ , $e=0.16$
at \RHIC\ gives a maximum non-uniform flow distribution.

\vskip0.5cm \cl{Table II \ \ The comparison of value of
model-parameters of net proton distributions}

\cl{ at AGS, SPS and RHIC} \vskip-0.5cm
\bcc\btbl{|c|c|c|c|c|}\hline
           & \multicolumn{4}{|c|}{Parameter}
        \\ \cline{2-5}
 energy region & $e$ & $y_{\rm e0}$ & $<\delta y>$ & $\chi^{2}/N$
\\ \hline
AGS & 0.7237 & 1.104 & 1.0 & 5.469
\\ \hline
SPS & 0.6535 & 2.105 & 2.0 & 3.003
\\ \hline
RHIC & 0.16 & 2.458 & $2.2\pm0.4$ & 4.134
\\ \hline \etbl\ecc

Therefore, the two parameters of $e$ and $y_{\rm e0}$ introduced
by \NUFM~ have their own physical meanings. $e$ is a physical
quantum that represents the non-uniformity of longitudinally
collective flow. From Table.II, we know that the ellipticity $e$
decreases with increasing energy from \AGS~ to \RHIC~. $y_{\rm
e0}$ represents the rapidity limit of emission source which
correspond to the nuclear stopping power, the larger of the value
of $y_{\rm e0}$ the smaller of the nuclear stopping power. As
shown in Fig.5 and Table.II, we can understand that at \RHIC~
energies collisions are quite transparent and the flow
distribution at longitudinal direction is more nonuniform.

\vskip0.8cm
\begin{center}{\bf C. Rapidity distributions of proton,
anti-proton and net-proton at \RHIC}\end{center}

\noindent The energy loss of colliding nuclei is a fundamental
quantity determining the energy available for particle production
(excitation) in heavy-ion collisions. Because the baryon number is
conserved, the rapidity distributions are only slightly affected
by re-scattering in later stage of the collision, the measured
net-baryon ($B-\overline{B}$)distribution retains information
about the energy loss and allows the degree of nuclear stopping to
be determined.

It can be seen from the Fig.7 that the \NUFM~ model reproduces the
central dip of the rapidity distribution of net-protons. given by
\BRAHMS~ Collaboration [5] working at \RHIC.

Fig.7 shows the resulting rapidity densities $dN/dy$ as a function
of rapidity. The most prominent feature of the distributions is
that proton and anti-proton show platter distributions,and
decrease at rapidities away from mid-rapidity . But the net-proton
distribution is absolutely different from that of the $p$ and
$\overline{p}$, it shows a dip at the central of rapidity
$y\simeq0$. From the distribution of $\overline{p}$ , we can study
the pair production of $p-\overline{p}$, which distributes nearly
uniformly in the longitudinal direction.

\vskip1.0cm
\begin{center}{\bf \large IV. Summary and Conclusions}\end{center}

\noindent In high energy heavy ion collisions, there will have a
large amount energy lose in relativistic heavy ion collisions, and
deposit at region of central rapidity. Soft hadronic observances
measure directly the final "freeze-out" stage of the collision,
when hadrons decouple from the bulk and free-stream to the
detectors. Freeze-out may correspond to a complex configuration in
the combined coordinate-momentum space, with collective components
(called flow) generating space-momentum correlations, as well as
geometrical and dynamical anisotropic. A detailed experimental
experimental-driven understanding of the freeze-out configuration
is the crucial first step in understanding the system's prior
evolution and the physics of hot colored matter. The isotropic
thermal model and cylindrical-symmetric longitudinal flow model
have discussed the collective flow feature,respectively. But they
cannot exhibit the central dip distribution especially at higher
collision energy when the nuclear have more transparency .
Non-uniform Flow Model(\NUFM) can not only discuss the collective
flow distribution feature, but also express the dip distribution.

In the paper we argue that the transparency/stopping of
relativistic heavy-ion collisions should be taken into account
more carefully. we assumes that the centers of the fireballs are
distributed non-uniformly in the longitudinal phase space. A
ellipticity $e$ is introduced through a geometrical
parametrization which can express the non-uniformity of flow in
the longitudinal direction,i.e., the centers of fireballs of
proton and net-proton prefer to accumulate in the two extreme
rapidity regions($y\approx\pm y_{e0}$) in the c.m.s. frame of
relativistic heavy ion collisions, and accordingly the
distribution is diluted in the central rapidity
region(($y\approx0$). It is found that the depth of the central
dip depends on the magnitude of the parameter $e$, i.e., the
non-uniformity of the longitudinal flow is describes the depth of
the central dip for proton,anti-proton and net-proton.

Comparing with our early work [8], we analyze and study the
features of collective flow and nuclear stopping at \RHIC, \SPS~
and \AGS~ energy regions are given in this paper,respectively.
Through comparing the feature of collision energy regions, it is
found that the uniformity of collective flow in the longitudinal
decrease with increasing collision energy, and much more
transparent with increasing incident energy. The feature of
net-proton distribution at \RHIC~ is completely different from
that of at \AGS~. Qualitatively speaking, the collision of
high-energy heavy-ions can be divided into two different energy
regions:the "baryon-free quark-gluon plasma region" (or the pure
quark-gluon plasma region)with $\sqrt{s}\geq 100$  GeV per
nucleon, and the baryon rich plasma region (or the stopping
region) with $\sqrt{s}\approx 5 {\sim} 10$ GeV per nucleon, which
corresponds to about many tens of GeV per projectile nucleon in
the laboratory system. In the baryon-free quark gluon plasma
region, we need to know the nuclear stopping power to determine
whether the beam baryons and the target baryons will recede away
from the center of mass without being completely stopped, leaving
behind quark-gluon plasma with very little baryon content. In the
baryon-rich region or the stopping region, the nuclear stopping
power determines whether the colliding baryons will be stopped in
the center-of-mass system and pile up to form a quark-gluon plasma
with a large baryon density.

This work was supported in part by the Major Science Foundation of
Education Department of Hubei Province of China Grant No. 2003Z002
and China Three Gorges University for Key Subjects Grant No.
2003C02.

\newpage
\noindent{\bf\large\bf Figure captions}

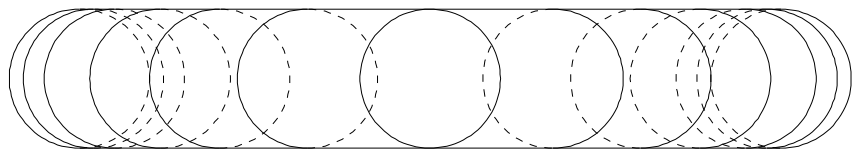
\vskip0.5cm \noindent  Fig.1 \ Schematic sketch of the
distribution of fire-balls in the non-uniform flow model (\NUFM).

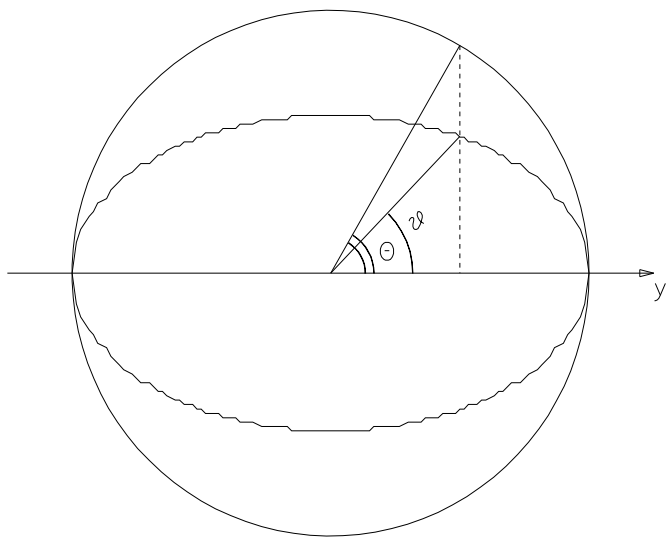
\vskip0.5cm \noindent  Fig.2 \ Schematic sketch of the emission
angle $\vartheta$ in the nonuniform flow model (\NUFM).

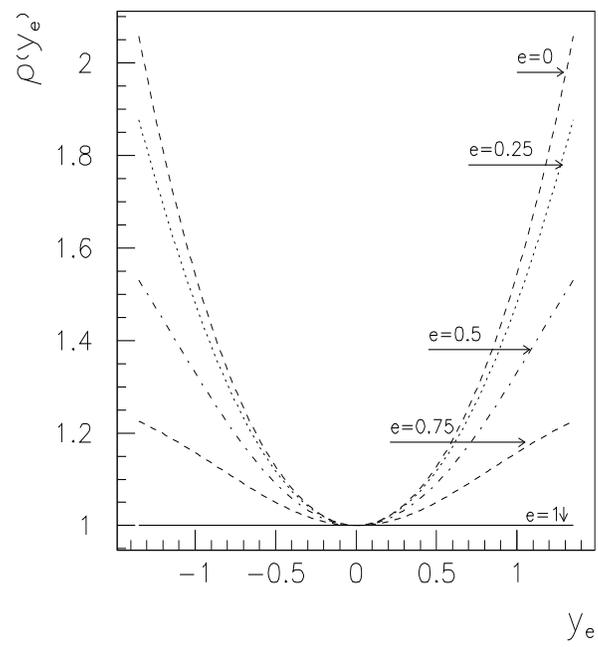
\vskip0.5cm \noindent  Fig.3 \ The distributions $\rho(y_{\rm e})$
of the center of fire-balls as a function of $y_{\rm e}$. When
$e\rightarrow 1$, the non-uniform distribution turns to the
uniform distribution $\rho(y_{\rm e})\rightarrow 1$.

\vskip0.5cm \noindent  Fig. 4($a,b,c$) are the rapidity
distributions for central 10.8, 8 and 6 GeV/$n$ Au+Au collisions
at \AGS, respectively. Open circles are the experimental data for
Au+Au collisions taken from Ref. [3]. dotted and solid lines are
the distributions from the isotropic thermal model, and
non-uniform longitudinal flow model (\NUFM) respectively.  The
temperature $T=0.12$ GeV.

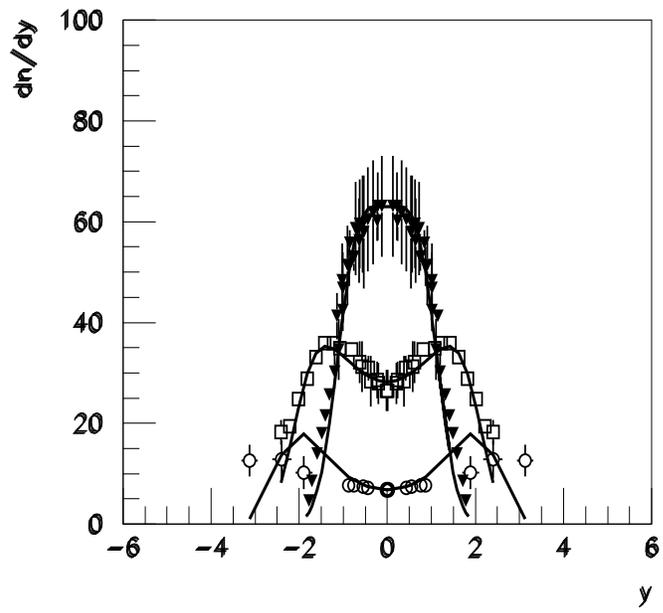
\vskip0.5cm \noindent  Fig.5 \ Rapidity distributions of
net-proton in central collisions at \AGS,\SPS~ and \RHIC. The
experimental data are taken from Ref.[5]. The solid line is our
calculation using the \NUFM\ model.  The temperature $T=0.12$ GeV.

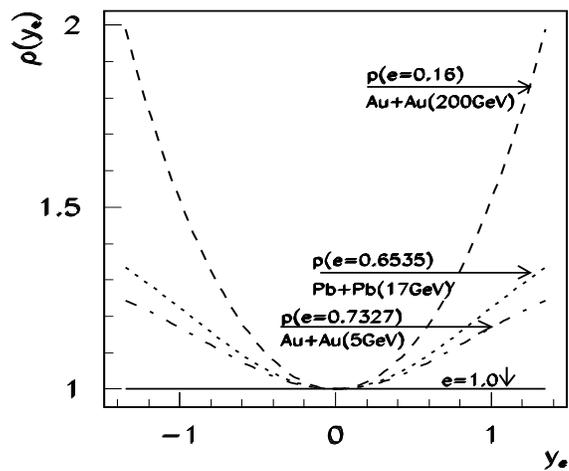
\vskip0.5cm \noindent Fig.6 \  The fire-ball distribution
functions $\rho(y_{\rm e})$ verus rapidity $y_{\rm e}$ with the
non-uniform flow model (\NUFM) at \AGS~,\SPS~ and \RHIC~ three
different energy regions.

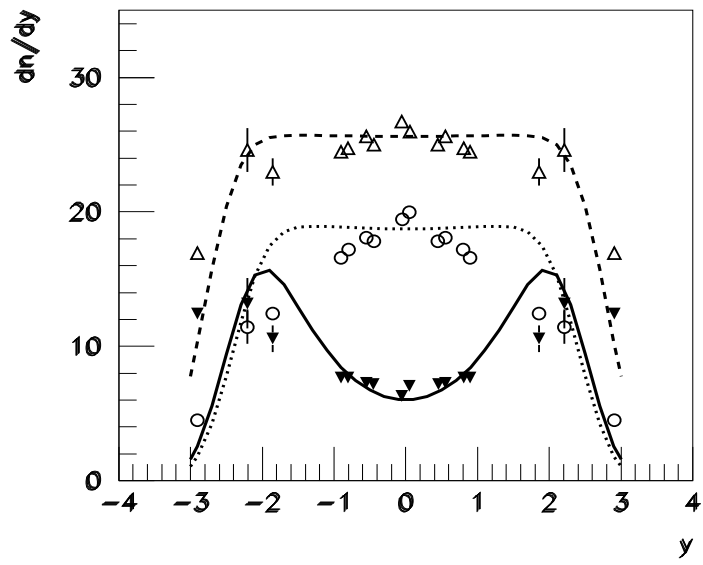
\vskip0.5cm \noindent Fig.7 \ Rapidity distributions of proton,
anti-proton and net-proton for central GeV/$n$ Au+Au collisions at
\RHIC. dotted line, dashed line and solid line are the calculation
results using \NUFM~ for proton, anti-proton and net-proton
distributions, respectively. The data are taken from Ref.[5].
(here, for proton: $e=0.9951$, $y_{eo}=3.2$; for anti-proton:
$e=0.9802$, $y_{eo}=2.4$; for net-proton: $e=0.16$, $y_{eo}=2.458$
.)

\newpage
\begin{center}
\begin{picture}(250,450)
\put(-40,380) { {\epsfig{file=fig1.epsi,width=320pt,height=54pt}}
} \put(0,20) { {\epsfig{file=fig2.epsi,width=250pt,height=200pt}}
}
\end{picture}
\end{center}

\vskip-12cm \cl{Fig. 1}

\vskip12cm \cl{Fig. 2}

\newpage
\begin{center} 
\begin{picture}(250,450)
\put(0,40) { {\epsfig{file=fig3.epsi,width=220pt,height=240pt}} }
\end{picture}
\end{center}
\vskip-1cm \cl{Fig. 3}

\newpage
\begin{center}
\begin{picture}(250,450) 
\put(-80,10) { {\epsfig{file=fig4.epsi,width=400pt,height=380pt}}
}
\end{picture}
\end{center}
\vskip-1.0cm \cl{Fig. 4}

\newpage
\begin{center} 
\begin{picture}(250,450)
\put(-80,100) { {\epsfig{file=fig5.epsi,width=350pt,height=300pt}}
}
\end{picture}
\end{center}
\vskip-4cm \cl{Fig. 5}

\newpage
\begin{center} 
\begin{picture}(250,450)
\put(-50,100) { {\epsfig{file=fig6.epsi,width=350pt,height=300pt}}
}
\end{picture}
\end{center}
\vskip-6cm \cl{Fig. 6}

\newpage
\begin{center} 
\begin{picture}(250,450)
\put(-100,100) {
{\epsfig{file=fig7.epsi,width=380pt,height=280pt}} }
\end{picture}
\end{center}
\vskip-4cm \cl{Fig. 7}

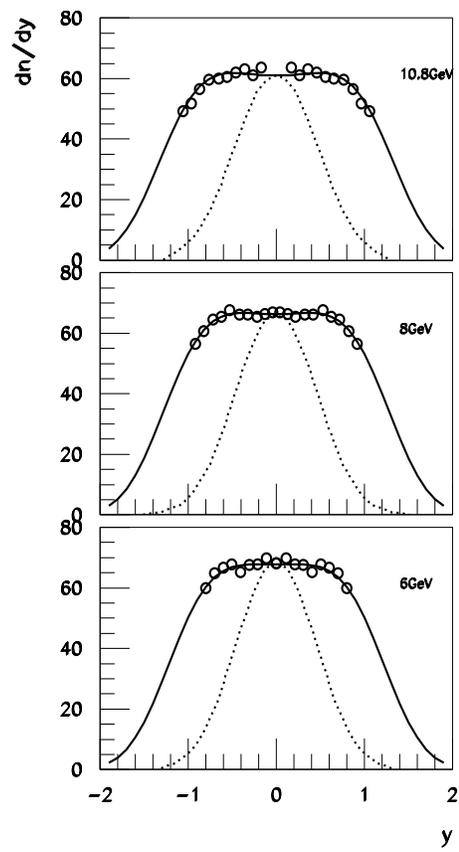
\end{document}